\documentclass[a4paper]{article}

\usepackage{fullpage} 
\usepackage{pstricks}
\usepackage{multicol}
\usepackage{graphicx}
\usepackage{movie15}
\usepackage{hyperref}
\usepackage[version=3]{mhchem}
\usepackage{amsmath} 
\usepackage{amssymb} 
\usepackage{amsfonts}
\usepackage{color}
\usepackage{soul}
\usepackage[margin=1.7cm,top=3cm,bottom=2.8cm,centering]{geometry}

\definecolor{rosepale}{rgb}{1.0, 0.7, 1.0}

\newcommand{\be}{\begin{equation}}
\newcommand{\ee}{\end{equation}}
\newcommand{\bea}{\begin{eqnarray}}
\newcommand{\eea}{\end{eqnarray}}

\title{Test of the formal basis of Arrhenius law with heat capacities\footnote{Published in Physica A. Stat. Mech. 510 (2018) 188-199}}
\date{} 

\author{Denis Michel$ ^{\ddagger} $ }

\begin{document}
\maketitle
\begin{small} $ ^{\ddagger} $ Universite de Rennes1-IRSET. Campus de Villejean. 35000 Rennes France. Email: denis.michel@live.fr. \end{small}
\\
\\
\begin{multicols}{2}
\noindent
\textbf{Abstract.} The exponential factor of Arrhenius satisfactorily quantifies the energetic restriction of chemical reactions but is still awaiting a rigorous basis. Assuming that the Arrhenius equation should be based on statistical mechanics and is probabilistic in nature, two structures for this equation are compared, depending on whether the reactant energies are viewed as the mean values of specific energy distributions or as particular levels in a global energy distribution. In the first version, the Arrhenius exponential factor would be a probability that depends once on temperature, while in the second it is a ratio of probabilities that depends twice on temperature. These concurrent equations are tested using experimental data for the isomerization of 2-butene. This comparison reveals the fundamental structure of the Arrhenius law in isothermal systems and overlooked properties resulting from the introduction of reactant energies into the equation.

\noindent
\textbf{Keywords}: Kinetics; equilibrium; heat capacity; enthalpy.\\

The laws of Arrhenius and van't Hoff are the basis of teaching and research in physical chemistry. Several versions of the energetic restriction of reactions have been proposed before the currently used Arrhenius equation, which works quite well but remains empirical \cite{Laidler}. A mathematical approach is used here to ground these equations on a rational basis. The probabilistic tool dedicated for this purpose is likely to be the geometric or exponential distribution, which proved sufficient to recover the Boltzmann statistics \cite{Michel2013,Michel2018}.

\section{Exponential distribution of energy}
In contrast to the historical equations of Arrhenius and van't Hoff, the new relationships between energy, kinetic and equilibrium constants described here are based on Boltzmann's laws of energy distribution. More precisely, the exponential distribution underlies both the Boltzmann theory and the thermochemical constants. Indeed, to define the distribution of maximum randomness in mechanical systems, mathematics provides a clear shortcut that bypasses Hamiltonians and Lagrangians: it is necessarily the "memoryless" exponential distribution \cite{Michel2013}. A probability distribution is based on a unique function called the probability density function (PDF). That of the exponential law is simply
\begin{equation} f(\mathcal{E})= \frac{1}{\left \langle \mathcal{E} \right \rangle}\ \large{\textup{e}}^{-\dfrac{\mathcal{E}}{\left \langle \mathcal{E} \right \rangle}} \end{equation} 
where $ \left \langle \mathcal{E} \right \rangle $ is the mean value of the distribution. We can apply it to energy distribution by assuming that $ \mathcal{E} $ is the number of energy units in a particle. This function, which decreases monotonically toward high energies, is illustrated in Fig.1, which shows the densities of particles containing different amounts of energy. The low energy particles are the most numerous and give the darker lower levels, while the particles become less numerous as the energy levels rise. Probabilities are obtained by integrating the PDF. The probability that a randomly chosen particle has an energy equal to or greater than a certain threshold $ \mathcal{E}^{\ddagger} $ is 

\begin{equation} P(\mathcal{E} \geqslant \mathcal{E}^{\ddagger})= \int_{\mathcal{E}=\mathcal{E}^{\ddagger}}^{\infty } f(\mathcal{E}) \ d\mathcal{E} =\large{\textup{e}}^{-\dfrac{\mathcal{E}^{\ddagger}}{\left \langle \mathcal{E} \right \rangle}} \end{equation} 

Note that the double dagger ($ \ddagger $), used here again, is historically used in rate theories to symbolize the energy threshold. To be validated by statistical mechanics, this approach should also allow to recover the Boltzmann distribution. This requirement is fully satisfied, since the probability that a particle has an energy level exactly of $\mathcal{E}^{\ddagger} $ can indeed be obtained by integrating the same PDF, but between $ \mathcal{E}^{\ddagger} $ and $ \mathcal{E}^{\ddagger}+1 $. 

\begin{equation} \begin{split} P(\mathcal{E} = \mathcal{E}^{\ddagger})&= \int_{\mathcal{E}=\mathcal{E}^{\ddagger}}^{\mathcal{E}^{\ddagger}+1} f(\mathcal{E}) \ d\mathcal{E} \\& = \left (1- \large{\textup{e}}^{-\dfrac{1}{\left \langle \mathcal{E} \right \rangle}}\right ) \large{\textup{e}}^{-\dfrac{\mathcal{E}^{\ddagger}}{\left \langle \mathcal{E} \right \rangle}}\\& =\dfrac{\large{\textup{e}}^{-\dfrac{\mathcal{E}^{\ddagger}}{\left \langle \mathcal{E} \right \rangle}}}{\sum \limits_{j=0}^{\infty }\large{\textup{e}}^{-\dfrac{j}{\left \langle \mathcal{E} \right \rangle}}} \end{split}  \end{equation} 
The latter form of Eq.(3) is known as the Boltzmann partition function, which is recovered here without the usual introductory treatment of statistical mechanics. Eq.(2), which is less familiar in thermochemistry, quantifies the probability for a particle to exceed a threshold energy. It could remind us of the Arrhenius formula

$$ k = A \ \large{\textup{e}}^{-\dfrac{E_{a}}{k_{B}T}} $$

but this is only apparent because the Arrhenius equation, as it is structured, cannot be a probability \cite{Michel2018}. To use Eq.(2) to construct the energy restriction formula for rate constants, we must first agree on the nature of the average energy $ \left \langle \mathcal{E} \right \rangle $. Identify $ \left \langle \mathcal{E} \right \rangle $ as $ k_{B}T $ allows one to use the exponential law described above to easily recover various extensions of statistical mechanics, including energy densities and Maxwell's velocity densities \cite{Michel2018}; but for an application to chemistry where different molecules have different energies, the nature of $ \left \langle \mathcal{E} \right \rangle $ should be precised. Two formulas are possible, depending on whether the interconvertible species belong to the same distribution or to two different distributions \cite{Michel2018}.

\subsection{Hypothesis of molecule type specific energy distributions}
Each type of molecule, with a specific name and defined by chemists on the basis of its covalent architecture, has its own specific energy, which could be thought of as its average energy. Within a homogeneous population of such a type of molecule, the energy of each individual molecule could take on different values due to thermal fluctuations. For example, a C=C double bond could be more or less rotated or stretched. This view is illustrated in Fig.1A, where $ E_{i} $ and $ E_{j} $ are the average energies corresponding to different values of $ \left \langle \mathcal{E} \right \rangle $ in equation (2), and individual molecules can have different energy levels. The probability that a given molecule has an energy higher than a threshold $ \mathcal{E}^{\ddagger} $ is exactly described by Eq.(2). Since the two interconvertible species have their own average energy, this hypothesis is called Model II. The extensions of this model are described in detail in \cite{Michel2018}.

\subsection{Hypothesis of the general mean energy}
In this second possibility illustrated in Figure 1B, all molecules, regardless of their chemical classification, participate in a global energy distribution characterized by a single average energy value. In this unique continuum of chemical energies, the covalent architecture used to classify molecules is no more important than secondary phenomena such as bond twisting and stretching. Because it is based on a single energy distribution, this hypothesis is called Model I.

\begin{center}
\includegraphics[width=8.2cm]{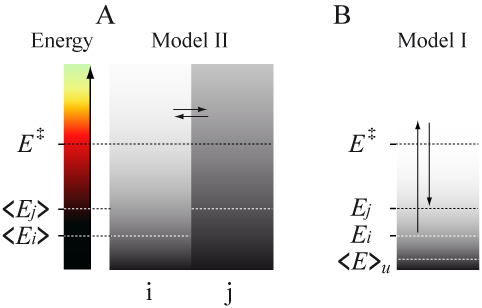} \\
\end{center}
\begin{small} \textbf{Figure 1}. Competing views of the energy restriction of rate constants. The quantities written in brackets are mean values of exponential distributions, whereas the quantities written without brackets are particular energy levels belonging to these distributions. (\textbf{A}) A mean energy value is assigned to each type of molecule $ i $ or $ j $. (\textbf{B}) The energies of the different types of molecules are not mean values but fixed values belonging to a common energy distribution with a general mean value $ \left \langle E \right \rangle_{u} $. \end{small}

\section{Applying these assumptions to the Arrhenius formula}

The two views of mean energies described above will be applied to the following generic form of the rate constant

\begin{subequations}  \label{E:gp}
\begin{equation} k = A_{(T)} \ \large{\textup{e}}^{B_{(T)}} \end{equation} 
which, like the Arrhenius equation, consists of a pre-exponential and an exponential term, both of which depend on the temperature. The pre-exponential factor $ A $ does not contribute to the reaction energy of the molecule, but corresponds to a configurational recurrence frequency, that is, the frequency with which one of the $ \varpi $ molecular configurations favorable to the reaction reappears out of a total number $ \omega $ of possible configurations.
\begin{equation} A_{(T)} =  \Phi T   \end{equation} 
\end{subequations}
with $ \Phi = \varpi k_{B}/(\Omega \ h) $, where $ k_{B} $ and $ h $ are the Boltzmann and Planck constants respectively. This pre-exponential factor will not be studied here, because the different theories more or less explicitly agree on its nature, including the activated complex theory \cite{Michel2018}. In contrast, the meaning of the Arrhenius exponent depends on the theory.

\subsection{Model II}

If the different types of reactants are assumed to have specific mean energies, the exponential term of Arrhenius would be the probability that a molecule of type $ j $ with mean energy $ \left \langle E_{j} \right \rangle $, reaches a threshold $ E^{\ddagger} $. This probability is predicted by the exponential distribution.

\begin{equation} k_{ji}/A_{ji}=P(E \geqslant E^{\ddagger})= \large{\textup{e}}^{-\dfrac{E^{\ddagger}}{\left \langle E_{j} \right \rangle}_{(T)}}  \end{equation}

The temperature dependence of this exponent is mediated by the mean energy $ \left \langle E_{j} \right \rangle_{(T)} $ only, because the threshold $ E^{\ddagger} $ is fixed and independent of temperature. As the temperature rises, $ \left \langle E_{j} \right \rangle_{(T)} $ increases while $ E^{\ddagger} $ remains unchanged, so the reaction rate is expected to increase. In this respect, this theory is in perfect agreement with the traditional interpretation of the stimulatory role of temperature on reaction rates through a shift of the entire Maxwell-Boltzmann distribution toward higher energies. As a consequence of this shift, the fractional particle population whose energy exceeds a certain threshold increases. Defining the equilibrium constant as a ratio of such rate constants, the van't Hoff equation becomes

\begin{equation} K_{ji}=\dfrac{A_{j}}{A_{i}} \ \large{\textup{e}}^{E_{ij}^{\ddagger } \left (\dfrac{1}{\left \langle E_{i} \right \rangle_{(T)}} - \dfrac{1}{\left \langle E_{j} \right \rangle_{(T)}} \right )} \end{equation}

which also depends on the temperature through the mean energies. 

\subsection{Model I}

In the unique chemical energy continuum of total mean $ \left \langle E \right \rangle_{u (T)} $ shown in Fig.1B, the reaction energies $ E_{i}(T) $ and $ E_{j}(T) $ and the threshold energy $ E^{\ddagger} $ are just particular energy levels. Given that the molecule has a basal energy $ E_{i} $, the Arrhenius equation can be defined as a conditional probability. The probability that the molecule has an energy higher than $ E^{\ddagger} $, given that its minimum energy is $ E_{i} $, is given by

\begin{subequations}
\begin{equation} k_{ij}/A_{ij} = P_{E \geqslant E_{i}}(E \geqslant E^{\ddagger}) \end{equation} 
which is
\begin{equation} \begin{split} P_{E \geqslant E_{i}}(E \geqslant E^{\ddagger})& = \dfrac{P(E \geqslant E_{i} \cap E \geqslant E^{\ddagger})}{P(E \geqslant E_{i})}\\ & = \dfrac{P_{E \geqslant E^{\ddagger}}(E \geqslant E_{i}) \ P(E \geqslant E^{\ddagger})}{P(E \geqslant E_{i})} \end{split} \end{equation}
and since obviously $ P_{E \geqslant E^{\ddagger}}(E \geqslant E_{i})=1 $

\begin{equation} k_{ij}/A_{ij}  = \dfrac{P(E \geqslant E^{\ddagger})}{P(E \geqslant E_{i})} =  \large{\textup{e}}^{-\dfrac{E^{\ddagger}-E_{i}}{\left \langle E \right \rangle_{u}}} \end{equation} 

At equilibrium, the concentration ratio between the interconvertible reactants, or the equilibrium constant, eliminates the energy threshold.
\begin{equation} \begin{split} K_{ji}&= \dfrac{A_{j}}{A_{i}} \ \dfrac{P(E \geqslant E^{\ddagger})}{P(E \geqslant E_{i})} \dfrac{P(E \geqslant E_{j})}{P(E \geqslant E^{\ddagger})}\\&= \dfrac{A_{j}}{A_{i}} \ \dfrac{P(E \geqslant E_{j})}{P(E \geqslant E_{i})}  \\&= \dfrac{A_{j}}{A_{i}} \ \large{\textup{e}}^{\dfrac{E_{j}-E_{i}}{\left \langle E \right \rangle_{u}}} \end{split} \end{equation}
\end{subequations}

In both hypotheses, the equilibrium distribution of the interconvertible molecules depends primarily on their energy difference, but under the hypothesis that the different molecules all have their own mean energy, this difference is further accentuated by the height of the energy barrier. The generalized chemical continuum model is consistent with the principle of thermochemical equilibrium in that the energy barrier does not interfere with the equilibrium. For this reason it appears simpler, but conversely its temperature dependence is more complex and this complexity is overlooked. In fact, the exponent of the rate constants depends twice on the temperature, contrary to what is suggested by the Arrhenius equation expressed as a straight line.

$$ B_{i (T)} =-\dfrac{E^{\ddagger}- E_{i (T)}}{\left \langle E \right \rangle_{u(T)}} $$
and that of the equilibrium constants depends three times on the temperature
$$ B_{i (T)} - B_{j (T)} = \dfrac{E_{i (T)}- E_{j (T)}}{\left \langle E \right \rangle_{u(T)}} $$

In this model, the access of a molecule to the energy threshold $ E^{\ddagger} $ is favored by an increase in temperature in two ways: (i) by increasing the general average population energy, which increases the denominator; and (ii) by increasing the specific energy of the considered reactant molecule, which decreases the numerator. The first mechanism is general and applies to all molecules in the isothermal medium, while the second mechanism depends on each type of molecule.

\subsection{Theoretical shapes of the Arrhenius plots in the two models}

The energy of the reactants, called enthalpy in thermochemistry, is rooted in statistical mechanics, but it results from the combination of many interfering mechanisms, which are particularly complex at low temperature. For simplicity, let us imagine a theoretical reactant consisting of a single resonator of frequency $ \nu $ (i.e. absorbing a single wavelength, contrary to the more advanced treatment of Debye), like an Einstein solid. In this respect, it may not be a coincidence that Einstein, in his original article on heat capacities, took as a model the diamond \cite{Einstein1907}, which contains only one type of chemical bond (C-C). The discrete approach to Planck's mean thermal energy can be quickly recovered from the number of ways to distribute $ E $ energy quanta into $ N $ particles.

\begin{equation} \Omega= \dfrac{(N+E-1)!}{(N-1)! \ E!} \end{equation}
giving a single particle mean entropy of

\begin{subequations} 
\begin{equation} \mathcal{S}= \dfrac{1}{N} \ln \Omega= \left (1+\frac{E}{N}  \right )\ln \left (1+\frac{E}{N}  \right ) - \dfrac{E}{N} \ln \left (\dfrac{E}{N}  \right ) \end{equation} 
The ratio $ E/N $ is equivalent to the mean number of energy quanta per particle $ U/h\nu $ \cite{Planck}. 
\begin{equation} \mathcal{S}= \dfrac{S}{k_{B}}= \left (1+\frac{U}{h\nu}  \right )\ln \left (1+\frac{U}{h\nu}  \right ) - \dfrac{U}{h\nu} \ln \left (\dfrac{U}{h\nu}  \right ) \end{equation}
\end{subequations}

Introducing temperature through the fundamental entropy equation

\begin{equation} \dfrac{dS}{dU}= \dfrac{1}{T} \end{equation}

yields 

\begin{subequations}
\begin{equation} \dfrac{k_{B}}{h\nu}\ln \left (1+\dfrac{h\nu}{U}  \right )  =\dfrac{1}{T} \end{equation} 
and
\begin{equation} U= \dfrac{h\nu }{\large{\textup{e}}^{h\nu / k_{B}T}-1} \end{equation} 
\end{subequations}

where $ h\nu $ can also be defined as $ h\nu =hc/\lambda $, where $ c $ is light velocity and $ \lambda $ is the spectroscopic wavelength of absorption. Hence, the number of energy quanta used in Eqs(1-3), is

$$ \left \langle \mathcal{E} \right \rangle=  \dfrac{U}{h\nu}=\dfrac{1}{\large{\textup{e}}^{h\nu / k_{B}T}-1} $$

Since the heat capacity of a substance is its ability to increase its energy with temperature, it can be derived from Eq.(11b), as did \cite{Einstein1907},

\begin{equation}  C=\dfrac{dU}{dT} = \dfrac{(h\nu)^{2}}{k_{B}T^{2}} \dfrac{\large{\textup{e}}^{h\nu/k_{B}T}}{\left(\large{\textup{e}}^{h\nu/k_{B}T}-1 \right )^{2}} \end{equation} 
The shape of this function is represented in Fig.2A, which shows that increasing either the temperature or the wavelength, symmetrically increases the heat capacity. The two models described above predict that the Arrhenius equation is not a straight line.

\subsubsection{Model I}

In the theory of the general mean energy, even without introducing the complexity of real heat capacities, the theoretical Arrhenius plot drawn for a single frequency is not linear. Taking $ k_{B}T  $ as the mean energy, Eq.(7c) becomes

\begin{equation}   \ln (k/A)= - \dfrac{E^{\ddagger}}{k_{B}T}+\dfrac{\dfrac{h\nu}{k_{B}T}}{\large{\textup{e}}^{\dfrac{h\nu}{k_{B}T}}-1} \end{equation}
represented in Fig.2B.

\subsubsection{Model II}

If neglecting the logarithmic temperature dependence of the preexponential factor, replacing the energy in Eq.(5) by that of Eq.(11b) gives,

\begin{equation}  \ln k/A = - \dfrac{E^{\ddagger}}{h\nu} \left(\large{\textup{e}}^{\dfrac{h\nu}{k_{B}T}}-1 \right ) \end{equation} 

whose Taylor series is
\begin{subequations} 
\begin{equation} \ln k/A= \dfrac{E^{\ddagger}}{h \nu} \left [1-\sum_{j=0}^{\infty} \left (\dfrac{h \nu }{k_{B} T} \right )^{j} / j!  \right ] \end{equation}
which can be limited to its first two terms when $ 1/T $ and $ \nu $ are small enough to render the higher order terms negligible
\begin{equation} \ln k/A= -\dfrac{E^{\ddagger}}{k_{B} T}-\dfrac{E^{\ddagger} h\nu}{2(k_{B}T)^2} \end{equation}
\end{subequations} 

This second term can generate convex Arrhenius plots for high frequency reactants (Fig.2C). It is interesting to note that the first term in Eq.(15b) is close to the traditional slope when $ H $ is negligible compared to $ E^{\ddagger} $. The traditional interpretation considers the Arrhenius slope to be constant, which misleadingly suggests that the Arrhenius plot is a straight line. We see that both theories predict that Arrhenius plots are globally curved. Moreover, these curvatures can be further complicated by irregularities in heat capacities, because these simplified plots hold for single resonators, whereas real molecules contain numerous adjacent resonators that absorb at different wavelengths and cooperatively contribute to the global energy of the molecule. Therefore, molecular energies cannot be determined from known wavelengths in a bottom-up approach, but can be deduced from the measured heat capacities in a reverse strategy.

\begin{center}
\includegraphics[width=8.5cm]{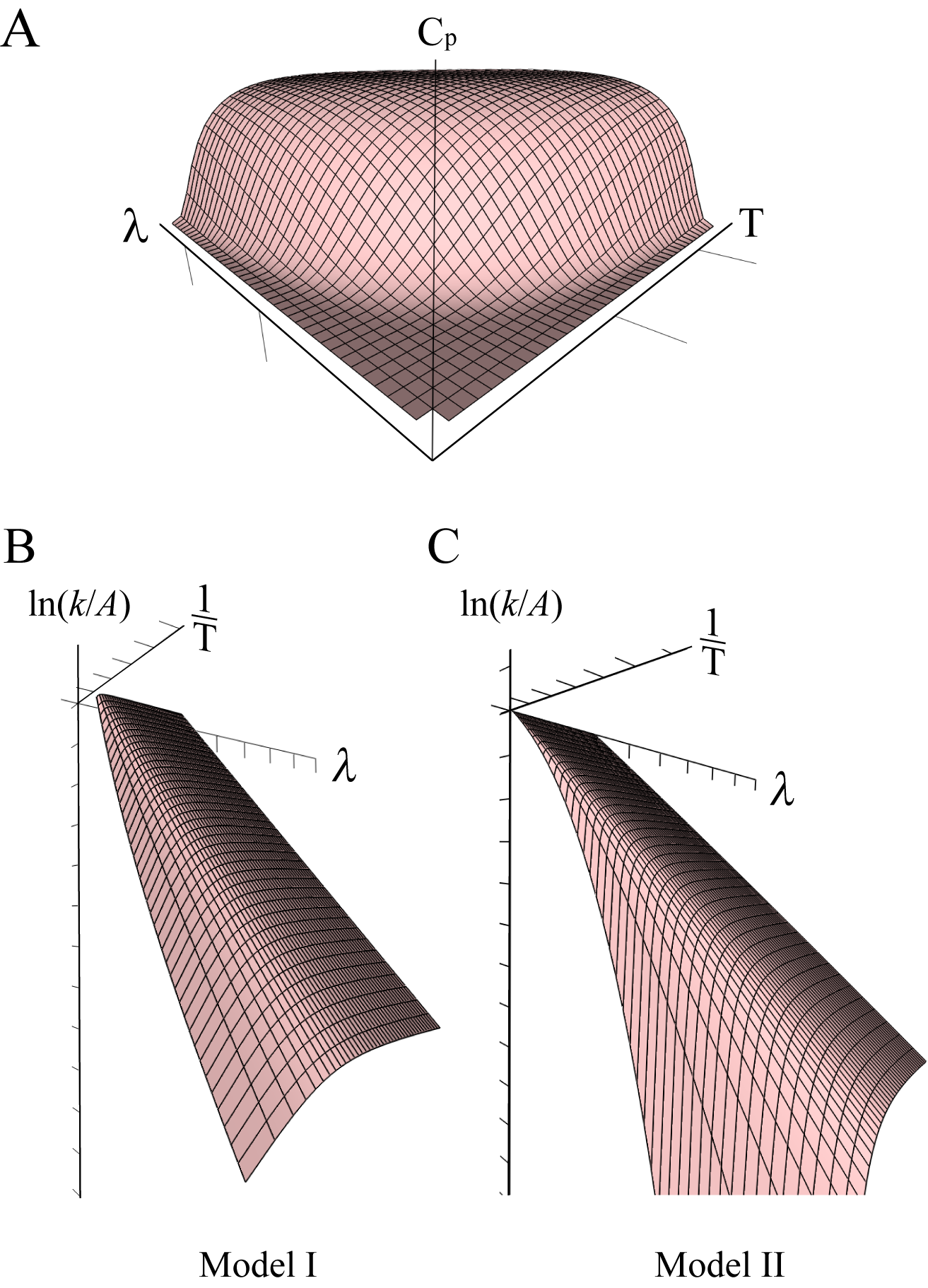} \\
\end{center}
\begin{small} \textbf{Figure 2}. Minimalist modeling of heat capacity and Arrhenius plots for a theoretical reactant made of a single resonator. (\textbf{A}) Heat capacity symmetrically increases when increasing either temperature or $ \lambda $. (\textbf{B}) Shape of the Arrhenius plot obtained from the energy continuum hypothesis (Eq.(13)), concave at high reactant frequency. (\textbf{C}) Shape of the Arrhenius plot with the hypothesis of a mean reactant energy (Eq.(14)), convex at high reactant frequency. The energy thresholds are arbitrary and the temperature dependence of the pre-exponential factor is not taken considered in these representations. \end{small}

\section{Experimental comparison of the approaches}
In physics, theories are not chosen for their elegance but for their capacity to describe experimental observations. The chemical reaction selected for this test is the isomerization between cis-2-butene ($ Z $) and trans-2-butene ($ E $), because it has a series of advantages: it involves reciprocal reactions unimolecular in both directions. Very precise data are available in kinetics \cite{Kaiser} and in equilibrium \cite{Kapteijn}. Moreover, the thermal isomerization of 2-butene is restricted by a quite high energy barrier (corresponding essentially to the energy needed to break the $ \pi $ bond), which should make it possible to clearly distinguish the two models described above, since the barrier plays a role in equilibrium in the former but not in the latter. Because of this barrier, the thermal conversion between the two isomers is not measurable at temperatures lower than 700 K. As a consequence, equilibrium data were obtained in presence of a catalyst \cite{Kapteijn}. Finally, the temperature-dependent heat capacities of these two isomers are also available, which will allow to approximate their energies.  

\section{Determination of absolute energies}
The two forms of the exponential factor compared here require knowledge of the absolute energies of the reactants. The probabilistic form based on reactant-specific average energies requires this value for the denominator of the exponent (Eq.(5)), while the form based on the energy continuum model requires it for the numerator of the exponent (Eq.(7c)). However, absolute energies and enthalpies are notoriously difficult to determine.

\subsection{What form of energy to use}
The energy of the single resonator reactant of frequency $ \nu $ used previously is theoretical and not available for the determination of the molecular energy of Eq.(11b), but the principle of this type of energy is retained here, that is, an energy equal to zero at $ T=0 $ K, increasing with temperature and calculable as the integral of the heat capacity, as theorized by Einstein \cite{Einstein1907} (Eq.(12)). This point is essential, since absolute energies are unknown, while heat capacities are often known, so we must first explicit their temperature dependence in the form of an integrable function. In the energy continuum model, similar to the current treatment of thermochemistry, absolute enthalpies ($ H $) are used, but enthalpies can take negative values, which would lead to a division by zero at a certain temperature in Eq.(8a), so only positive energies (written $ E $) are used for Model II.

\begin{center}
\includegraphics[width=8cm]{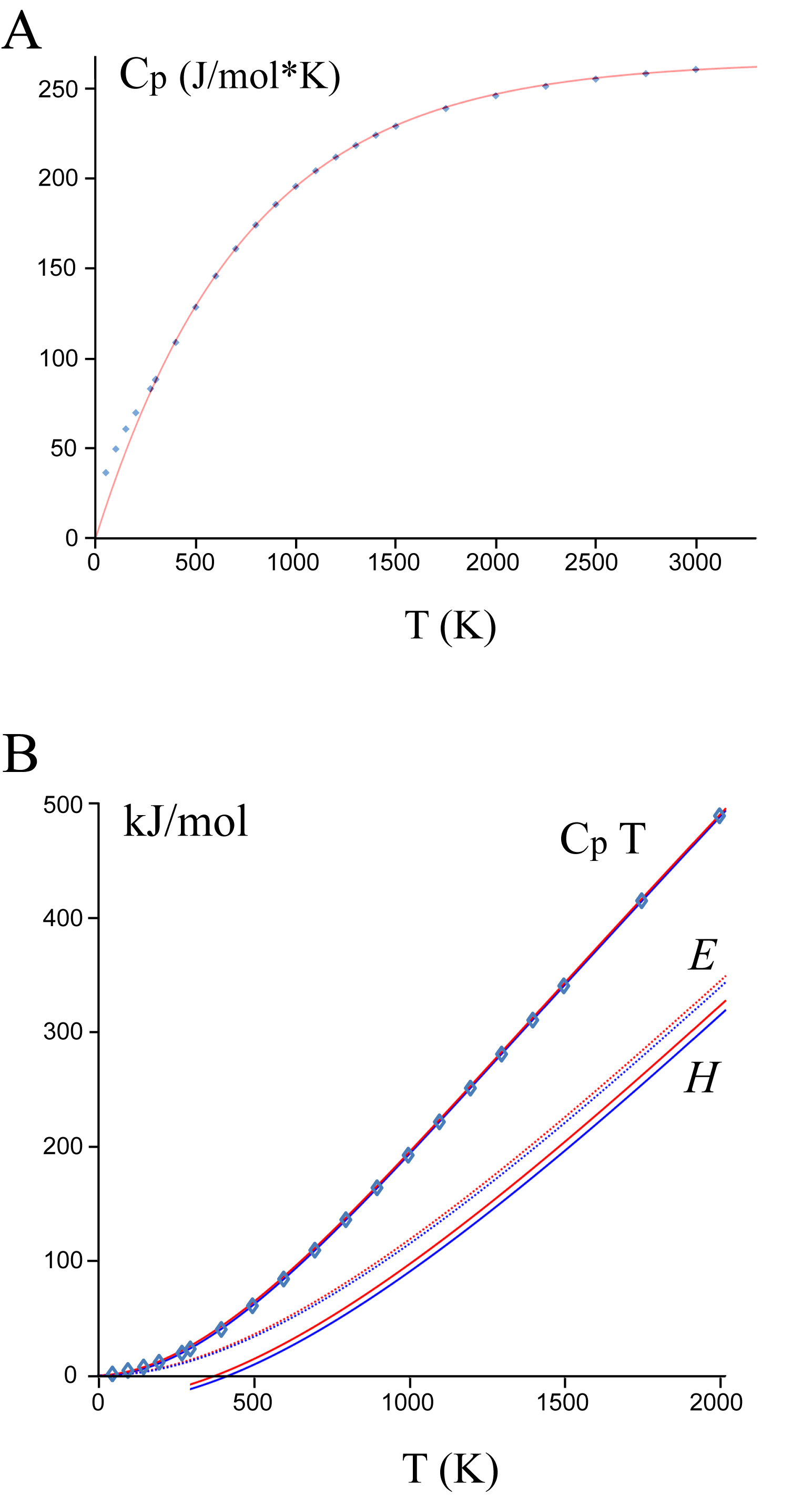} \\
\end{center}
\begin{small} \textbf{Figure 3}. (\textbf{A}) Measured heat capacities and fitting function valid at high temperatures. (\textbf{B}) Absolute enthalpies ($ H $, solid lines) and positive energies ($ E $, dotted lines) of the 2-butene isomers, red for the cis isomer and blue for the trans isomer. For comparison, energies calculated as direct products $ T C_{p}(T) $ are shown (diamonds correspond to trans-2-butene).\\ \end{small}

Absolute enthalpies can in principle be obtained from heat capacities using the Kirchhoff approach

\begin{equation}  H(T)=H(T_{0})+\int_{T_{0}}^{T}C_{p}(T) \  dT \end{equation} 

where $ T_{0} $ is temperature of reference chosen for convenience.
The first step is to determine a function satisfactorily adjusted to the heat capacities measured by the \textit{Thermodynamics Research Center} \cite{TRC} and available in the NIST Chemistry WebBook. The 3 parameter exponential association 

\begin{equation} C_{p} \approx \alpha \left(\beta-\large{\textup{e}}^{- \gamma T} \right ) \end{equation} 

is chosen here as a fitting function because it is easily integrable and valid for a wide temperature range over 300 K. The fitting parameters are, for the cis isomer $ \alpha=264.49 $, $ \beta= 1 $ and $ \gamma = 13.35 \times 10^{-4} $ (curve shown in Fig.3A), and for the trans isomer $ \alpha= 266.07 $, $ \beta= 0.9968  $ and $ \gamma = 12.97 \times 10^{-4} $.\\

The indefinite integrals $ F(T) $ of these functions can then be used to calculate absolute enthalpies provided the enthalpy of formation is known at a given temperature $ T_{0} $.

\begin{equation} H(T)=F(T)-F(T_{0})+ H_{f}(T_{0}) \end{equation} 

Taking the starting elements as 0 J/mol, the values given in the NIST WebBook are for the cis isomer $ H_{f Z}(298) $= -7.7 kJ/mol and for the trans isomer $ H_{f E}(298) $= -10.8 kJ/mol. \cite{Prosen}. To relate these values to the validity range of the exponential fitting curve, the intermediate range is covered by a polynomial approximation (Appendix). The enthalpies obtained in this way are shown in Fig.3A. For comparison, the values $ T C_{p}(T) $ are plotted in the same graph (Fig.3B) to show that this direct approach overestimates the energies.

Based on the view of molecular energy as a positive number of energy quanta, as in Eq.(11b), and to avoid division by zero in the probabilistic approach, internal energies are simply defined as

\begin{equation} E(T)=F(T)-F(0) \end{equation} 

The models can now be compared numerically with the kinetic results.

\section{Determination of energy threshold from the Arrhenius slope at $ T $= 800 K}
The rate equations of cis trans isomerization of 2-butene, determined experimentally \cite{Kaiser} are 
$$ k_{Z \rightarrow E}= 9 \times 10^{13} \times \textup{Exp}\left(-\dfrac{31845}{T}\right) \ s^{-1} $$
and
$$ k_{Z \rightarrow E}= 21 \times 10^{13} \times \textup{Exp}\left(-\dfrac{32725}{T} \right) \ s^{-1} $$

The slopes are expressed in Kelvin units to eliminate any model assumptions. Written in this form, these equations are purely experimental and model independent, so they can be used as templates to calculate the parameters of the model derived equations. Using the central experimental temperature of 800 K as the best confidence level, the parameters are adjusted so that at this temperature the slope $ d(\ln k)/d(1/T) $ of the theoretical equations gives the experimental slope. Superimposing the resulting curves on the Arrhenius plots of \cite{Kaiser} confirms the concavity of model I (Fig.4A) and the convexity of model II (Fig.4B), both predicted theoretically (Fig.2). Introducing energies into the Arrhenius equation should yield the same value for $ E^{\ddagger } $ by two different calculations based on the two reciprocal reactions.

\subsection{Model I}

The two types of energy will be applied to this model. 

\subsubsection{Using absolute enthalpies}

We find, for the cis isomer,
\begin{itemize}
\item $ A_{Z}=2.82 \times 10^{4} \ s^{-1} $
\item $ E^{\ddagger }= 185.85 $ kJ/mol
\end{itemize}

and for the trans isomer,
\begin{itemize}
\item $ A_{E}=25.00 \times 10^{4} \ s^{-1} $
\item $ E^{\ddagger }= 190.83 $ kJ/mol
\end{itemize}

The pre-exponential factors, defined as the intersection of the Arrhenius line with the vertical axis, are of course no longer relevant in the two models using reactant energies. The values reported in \cite{Kaiser} were obtained using the linear Arrhenius equation, dependent on temperature only through $ RT $, and without taking into account the contribution of enthalpies. Since the experiments giving these results were performed at temperatures around 800 K (1/$ T $ =0.00125) \cite{Kaiser}, the Arrhenius slopes obtained under these conditions must be compatible with the reactant enthalpies at this temperature, which, using the Kirchoff approach described above, are $ E_{Z}(800)=60 $ kJ/mol and $ E_{E}(800)=54 $ kJ/mol.

The values of $ E^{\ddagger } $ found from the two reciprocal reactions, 186 and 191 kJ/mol, are not very distant, but the activation energies however, are no longer those currently reported. The previous values were $ E_{a Z} $= 264.6 kJ/mol and $ E_{a E} $= 271.9 kJ/mol. They are now $ E_{a Z} $= 125.2 kJ/mol and $ E_{a E} $= 136.3 kJ/mol at 800 K.
 
\subsubsection{Using positive energies} 
 
The same treatment with positive energies gives very similar results

\begin{itemize}
\item $ A_{Z}=2.82 \times 10^{4} \ s^{-1} $
\item $ E^{\ddagger }= 207.34 $ kJ/mol
\end{itemize}

and for the trans isomer,
\begin{itemize}
\item $ A_{E}=27.00 \times 10^{4} \ s^{-1} $
\item $ E^{\ddagger }= 214.28 $ kJ/mol
\end{itemize}
 
\subsection{Model II}

Under the hypothesis of reactant-specific energy distribution and positive energies, since the reactant energies are much higher than $ RT $, the values of $ E^{\ddagger } $ will be much higher than the previous ones. Moreover, since the calculations from the reciprocal reactions use multiplications, the determination of $ E^{\ddagger } $ is expected to be much more sensitive to inaccuracies than the previous one based on subtractions. Incorporating the positive energies calculated above into the probabilistic rate constant of Eq.(5) and adjusting the slope of the plot in Arrhenius coordinates at $ T $= 800 K, one finds for the cis isomer,

\begin{itemize}
\item $ A_{Z}=2.43 \times 10^{6} \ s^{-1} $
\item $ E^{\ddagger }= 1906.53 $ kJ/mol
\end{itemize}

and for the trans isomer,
\begin{itemize}
\item $ A_{E}=6.25 \times 10^{6} \ s^{-1} $
\item $ E^{\ddagger }= 1849.28 $ kJ/mol
\end{itemize} 

Thus, the kinetic data give different parameter values depending on the model, without providing a discriminating criterion. The comparison of kinetic and equilibrium data will provide this discrimination.

\section{Comparison of the data obtained from Arrhenius coordinates with the results obtained at equilibrium}
The results of \cite{Kapteijn} obtained with catalysis are remarkably precise and compatible with earlier, more partial measurements, as for example \cite{Scott}. 

\begin{center}
\includegraphics[width=7.2cm]{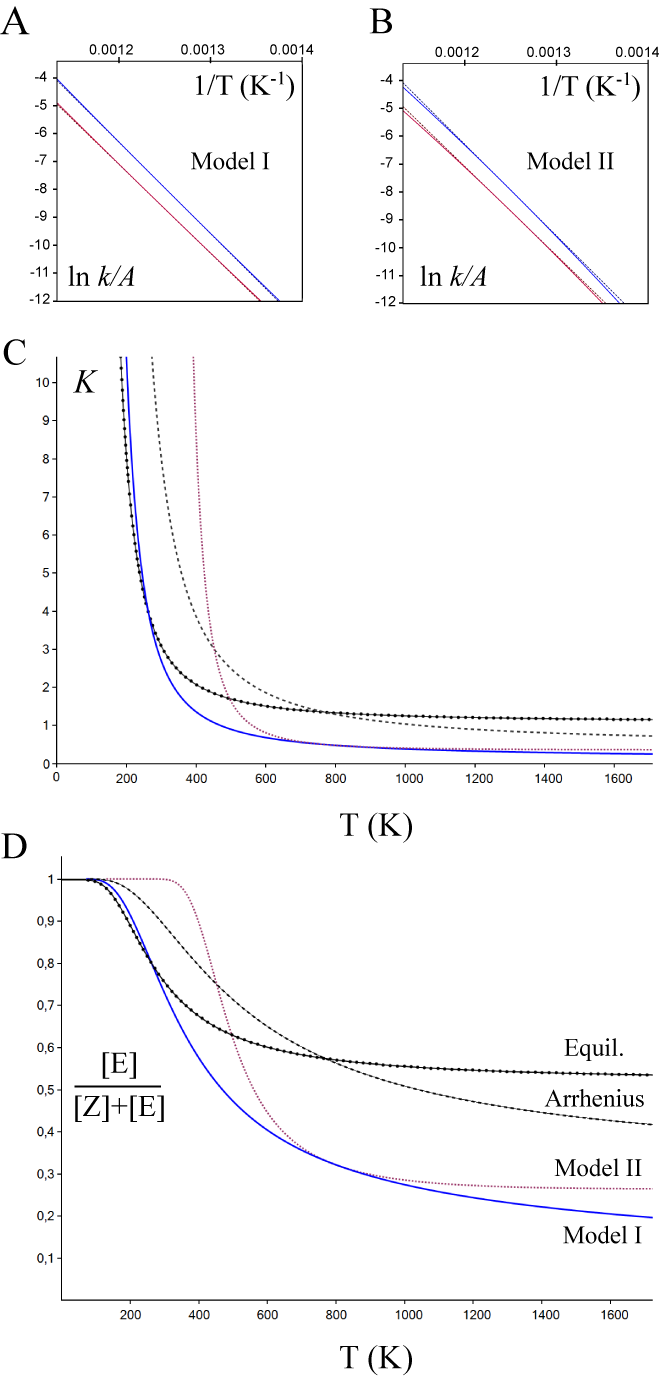} \\
\end{center}
\begin{small} \textbf{Figure 4}. (\textbf{A}) Arrhenius plots of the reactions of 2-butene isomerization reported in \cite{Kaiser} (dashed lines), superimposed to those obtained by introducing the reactant energies into the Arrhenius equation according to Eq.(7c). Cis to trans reaction in blue and trans to cis reaction in red. Theoretical parameters were obtained by adjusting the slopes at 800 K. (\textbf{B}) Corresponding results when reactant energies are introduced through Eq.(5). (\textbf{C}) Comparison of the kinetic results obtained in Arrhenius coordinates \cite{Kaiser} with the equilibrium constant \cite{Kapteijn}. (\textbf{D}) Proportion of the trans isomer in equilibrium predicted by the different theories. The curves shown in (C) and (D) are (i) the equilibrium data of \cite{Kapteijn} (black line with dots) in which the cis isoform is always predominant at all temperatures tested; (ii) the ratio of Arrhenius equations of rate constants $ k_{Z \rightarrow E}/ k_{E \rightarrow Z} $ of \cite{Kaiser} (black dashed line, for which the switch between $ Z $ and $ E $ isomers occurs at 1000 K); the curve predicted in the Model I (blue plain line, for which the equivalence between $ Z $ and $ E $ isomers is obtained at 468 K); the curve predicted using the Model II (dotted line, equivalence between $ Z $ and $ E $ at 560 K).\\ \end{small}

Fig.4C and D show the comparison between the values of the equilibrium constants $ K_{Z \rightarrow E} $ obtained in \cite{Kapteijn} and the ratios of kinetic constants $ k_{Z \rightarrow E}/k_{E \rightarrow Z} $ based on the values measured in \cite{Kaiser} and interpreted as either (i) the linear Arrhenius equations reported in \cite{Kaiser}, (ii) the Arrhenius equations modified to include reactant energies (Model I) and (ii) the probabilistic equation where the reactant energies are envisioned as mean values (Model II). All approaches confirm the predominance of the trans-isomer at low temperatures and the accumulation of the cis-isomer at higher temperatures, but none gives exactly the equilibrium results. Obviously and even in the case of uncertainties on the values of the energies, the model II is disqualified since it predicts that cis-2-butene is almost absent below 300 K, in contrast to the experimental measurements \cite{Scott,Kapteijn}, in which this isoform is clearly present from 100 K. This difference is specific of the model II because it is due to the participation of the activation thresholds in equilibrium. As a matter of fact, using the kinetic model II, the presence of trans-2 butene in equilibrium can be shifted at lower temperatures by decreasing $ E^{\ddagger } $ in the equations (not shown).

\section{Conclusions}
The comparison between kinetic and equilibrium experimental data shows that the hypothesis that the Arrhenius exponential factor could be a probability does not hold for a homogeneous temperature chemical system. This model can be applied to other physical situations such as two-temperature non-equilibrium systems or the evaporation of a lump of hot matter \cite{Michel2013}, but not to thermochemistry. The basic principle that activation energies do not affect equilibrium conditions is verified, but it is very instructive to return to the theoretical foundations to see what this principle implies. This reductionist approach sheds light on (i) the nature of enthalpies, which are not average energies but fixed initial values that can be increased by thermal fluctuations, and (ii) the nature of the Arrhenius equation, which is a ratio of exponential probabilities calculated within a single energy distribution. This universal energy continuum encompasses all types of molecules, regardless of our classification of the chemical world into different molecular species. Covalent bonds, which are used to define molecules, are no more important in defining molecular energy than other features such as non-covalent bonds, stretching, rotation, and vibration. In fact, this view of molecular energy is consistent with its immaterial nature, homogenizing between colliding molecules without regard to their structure. In this global exponential distribution, all forms are metastable and the most energetic molecules are the rarest.

The acceleration of reactions by temperature is mediated on two levels: by increasing the denominator of the exponent $ k_{B}T $ and by decreasing the numerator by reducing the difference $ E_{a}=E^{\ddagger}-H (T) $. Curiously, this latter role of temperature is generally ignored in the interpretation of the Arrhenius equation as a straight line. The usual explanation of the role of temperature on reaction rates is based on a shift toward the high energies of a Maxwell-Boltzmann distribution, which increases the fraction of molecules exceeding a fixed energy threshold. Strangely enough, this pedagogical interpretation is fully valid for the reactant-specific mean energy hypothesis rejected here, but it is very partial for the universal mean energy hypothesis, since it does not take into account the enthalpy of the reactants. The introduction of the reactant energy in the Arrhenius equation and the replacement of the activation energies by energy thresholds common to the two reciprocal reactions would allow to anticipate the non-linearity of the Arrhenius plots. The simple knowledge of heat capacities allows to estimate this energy and to predict many kinetic and thermodynamic behaviors. \\

\noindent
\textbf{Acknowledgements}. I thank very much Bill Kaiser and Freek Kapteijn for helpful discussions.

\newpage
\begin{center}
\Huge{Appendices}
\end{center}

\appendix
\setcounter{equation}{0}  
\numberwithin{equation}{section}

\section{Assembly of single resonator heat capacities}
Real molecular heat capacities differ from Einstein's theory in that they are not derived from a single absorption wavelength. A first attempt to construct the global heat capacity of a molecule can be done by juxtaposing its internal resonators. The so-called Neumann-Kopp law states that the heat capacity of a composite material is equal to the weighted average of the individual elemental heat capacities. Using this postulate, the measured heat capacities can be fitted to a theoretical curve of the form

\begin{equation}  f(T)=A+ \dfrac{B}{n}\sum_{\lambda_{1}}^{\lambda_{n}}\mathcal{C}(\lambda_{j}, T) \end{equation} 
where $ \mathcal{C}(\lambda_{j}, T) $ is the normalized shape of the Einstein capacity for the chosen wavelength $ \lambda_{j} $, ranging from 0 to 1 when either $ \lambda $ and/or $ T $ go from 0 to infinity.

\begin{equation} \mathcal{C}(\lambda_{j}, T)=  \left (\dfrac{hc}{k_{B}T\lambda_{j}}  \right )^{2} \dfrac{\large{\textup{e}}^{\frac{hc}{k_{B}T\lambda_{j}}}}{(\large{\textup{e}}^{\frac{hc}{k_{B}T\lambda_{j}}}-1)^{2}} \end{equation}

and $ A $, $ B $ are constants that are used to adjust the scale of $ \mathcal{C} $ to the actual values of $  C_{p} $, in particular by taking into account the non-negligible values of the heat capacities measured near $ T=0 $. Using the infrared absorption wavelengths reported for the two isoforms of 2-butene in the NIST Chemical WebBook, this averaging approach yields

\begin{equation} C_{E}= 61+213 \ [\mathcal{C}(3.5 \mu \textup{m})+ \mathcal{C}(7 \mu \textup{m})+\mathcal{C}(10.5 \mu \textup{m})]/3 \end{equation}
and

\begin{equation} C_{Z}= 18+255 \ [\mathcal{C}(3.5 \mu \textup{m})+ \mathcal{C}(7 \mu \textup{m})+\mathcal{C}(10.5 \mu \textup{m})+\mathcal{C}(16 \mu \textup{m})]/4 \end{equation}

This equation is satisfactory for trans-2-butene, but less successful for cis-2-butene (Fig.A1C). Although this isoform has two additional absorption wavelengths compared to its cis counterpart (about 15 and 17 $ \mu m $), the best result was actually obtained using only one wavelength, with the intermediate value of 16 $ \mu m $. This method, however, yields long and hardly tractable equations. In addition, it is only moderately justified because a molecule cannot be approximated as a juxtaposition of resonators, since they are not independent and are likely to cooperate. The different peaks of the infrared absorption spectrum may be involved in different molecular phenomena, including C-H, C-C, and C=C bond stretching or rocking, which are not independent of each other. Therefore, it may be preferable to use a more arbitrary but simpler formula, such as the simple exponential function used in the main text and the polynomial function shown below, which is useful for intermediate temperatures.

\section{Polynomial fitting function for heat capacities between 273 and 1500 K}

\begin{equation} C_{p} \approx \alpha +\beta \ T+\gamma \ T^{2}+\delta \ T^{3}+\epsilon \ T^{4}   \end{equation} 
For the cis isomer, $ \alpha=7.63 $, $ \beta= 31.35 \times 10^{-2} $, $ \gamma=-1.63 \times 10^{-4} $, $ \delta=4.08 \times 10^{-8} $ and $ \epsilon=-4 \times 10^{-12} $. For the trans isomer, $ \alpha=-6.21 $, $ \beta= 34.13 \times 10^{-2}  $, $ \gamma=-1.86 \times 10^{-4} $, $\delta =4.91 \times 10^{-8} $ and $ \epsilon=-5.07 \times 10^{-12} $.\\

\end{multicols}

\begin{center}
\includegraphics[width=17cm]{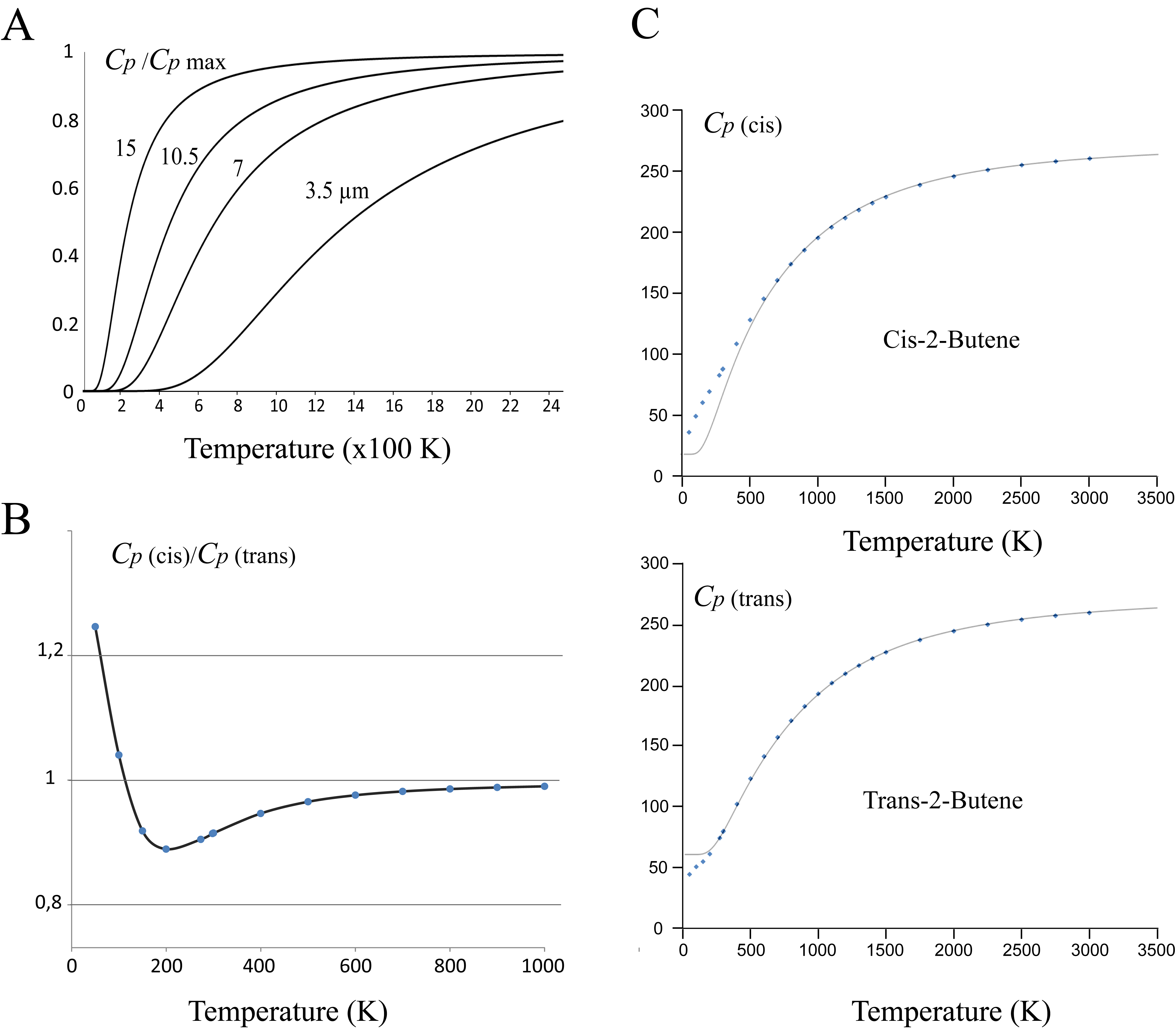} \\
\end{center}
\begin{small} \textbf{Figure A1}. Heat capacities of the cis and trans isomers of 2-butene. (\textbf{A}) Normalized Einstein heat capacities (Eq.(A.2)) for the absorbed infrared wavelengths of cis-2-butene taken individually. (\textbf{B}) The complexity of the real heat capacities is particularly evident at low temperatures, where the relative values obtained for the two isomers of 2-butene are surprisingly different. (\textbf{C}) Fit between the measured heat capacities, and functions defined as the sum of the elementary heat capacities. The fit is good at high temperatures but not at low temperatures. \end{small}

\end{document}